\documentstyle[12pt]{article} % usual 12pt
\catcode`\@=11
\def	\eqnum		#1{(\ref{#1})}       %equation # in round parenthesis
	\newdimen\eqskip
	\newdimen\txtskip
	\baselineskip=\txtskip
\def    \be             {\begin{equation}}
\def    \ee             {\end{equation}}
\def    \ba             {\begin{eqnarray}}
\def    \ea             {\end{eqnarray}}

\def    \=              {\;=\;}
\def    \frac           #1#2{{#1 \over #2}}
\def    \ret            {\\[\eqskip]}

\def \chiz {\mbox{$\chi_{c0}$}}
\def \chio {\mbox{$\chi_{c1}$}}
\def \chit {\mbox{$\chi_{c2}$}}
\def \as   {\mbox{$\alpha_s$}}
\def \assq {\mbox{$\alpha_s^2$}}

\def \asopi {\frac{\alpha_s}{\pi}}
\def \oacube {\mbox{${\cal O}(\alpha_s^3)$}}

\def \pt   {\mbox{$p_t$}}
\def \mc   {\mbox{$m_c$}}
\def \mcfour {\mbox{$m_c^4$}}
\def \ebin {\mbox{$E_{bind}$}}
\def \aem  {\mbox{$\alpha_{em}$}}
\def \aemsq{\mbox{$\alpha_{em}^2$}}
\def \to   {\mbox{$\rightarrow$}}
\def \rprime {\mbox{$\vert R^{\prime}(0) \vert^2$}}
\def \hone {\mbox{$H_1$}}
\def \height {\mbox{$H_8$}}
\def \chizgg {\mbox{$\Gamma(\chiz \to \gamma\gamma)$}}
\def \chitgg {\mbox{$\Gamma(\chit \to \gamma\gamma)$}}
\def \chijgg {\mbox{$\Gamma(\chi_J \to \gamma\gamma)$}}
\def \chizlh {\mbox{$\Gamma(\chiz \to LH)$}}
\def \chiolh {\mbox{$\Gamma(\chio \to LH)$}}
\def \chitlh {\mbox{$\Gamma(\chit \to LH)$}}
\def \chijlh {\mbox{$\Gamma(\chi_J \to LH)$}}
\def \GeV    {\mbox{GeV}}
\def \MeV    {\mbox{MeV}}

\begin{document}
\begin{titlepage}
\nopagebreak
{\flushright{
        \begin{minipage}{5cm}
        CERN-TH/95-67 \\
	arch-ive/9503465 \\
        \end{minipage}        }

}
\vfill
\begin{center}
{\LARGE { \bf \sc An update on $\chi_c$ Decays: \\[0.5cm]
 Perturbative QCD versus Data} }
\vfill
\vskip .5cm
{\bf Michelangelo L. MANGANO,}
\footnote{On leave of absence from INFN, Pisa, Italy}
\vskip .3cm
{CERN, TH Division, Geneva, Switzerland} \\
\vskip .6cm
{\bf Andrea PETRELLI}
\vskip .3cm
Dipartimento di Fisica dell'~Universit\`a and INFN, \\
Pisa, Italy
\end{center}
\nopagebreak
\vfill
%\vskip 3cm
\begin{abstract}
We present a global fit of current available experimental results on $\chi_c$
decays within next-to-leading-order perturbative QCD.  The quality and reduced
errors of recent data improve the agreement between theory and
experiment.
\end{abstract}
\vskip 1cm
CERN-TH/95-67 \hfill \\
March 1995 \hfill
\vfill
\end{titlepage}

The study of charmonium has recently received renewed interest. All aspects,
from spectroscopy, to decays and production have benefited from new data
and novel theoretical developments. Recent experimental data concern
measurements of total \cite{E760-tot,Cester}\ and radiative decay widths
\cite{E760-gg,TPC2-gg,CLEO-gg},
the observation of the
$^1P_1$ state \cite{E760-h}, measurements of decay branching ratios
of B mesons into P-wave
charmonium states \cite{CLEO-btochi},
production cross sections in hadronic collisions, both
at fixed target \cite{fxt} and at collider energies
\cite{UA1-psi,CDF-psi,D0-psi}.
New theoretical developments cover,
among other things, attempts to extract the value of \as\ from lattice QCD
calculations of quarkonium decay widths \cite{mack},
the development of a systematic
approach to the problem of infrared ambiguities in the production and decay of
P-wave states \cite{bbl1,bbl2},
discovery of new production mechanisms at high \pt\ \cite{highpt}. The large
disagreement observed between expected and measured production rates at the
Tevatron \cite{CDF-psi}, has also led to new speculations about the existence
of exotic spectroscopy in the charmonium system \cite{close}.
We believe that this discrepancy
provides an important arena in which to test our understanding of the boundary
domain between perturbative and non-perturbative QCD. In this respect, it is
fundamental to derive the best possible perturbative predictions, in order to
be able to firmly assess the need for, and to properly model, possible
additional non-perturbative phenomena required for a complete description of
this physics. With this goal in mind, we recently completed a full
next-to-leading-order (NLO)
calculation for the total production cross section of $^3P_{0,2}$ states in
hadronic collisions. A consistent use of this calculation, requires the
inclusion of phenomenological parameters extracted from a NLO analysis of
inclusive decay widths of these states. In this letter, we therefore present an
updated comparison between QCD predictions for the decay widths of $^3P_J$
($\chi_{cJ}$) states and the latest experimental data. We will not include the
$^1P_1$ state here, as the available data are still insufficient.

Several detailed studies of this subject have appeared in the past
\cite{rosner,bbl1,consoli,schuler}. We feel that the
new data justify an update. Furthermore, we improve the commonly used
expression for the decay width of the $\chi_{c1}$ state to light hadrons
\cite{barbieri}, with the inclusion of a finite \oacube\ term which has
usually been neglected.

We will show that the inclusion of all available data, in addition to relaxing
some theory constraints used in the literature so far, allows a consistent
global fit in terms of three parameters. The first parameter is
simply \as, the QCD coupling constant. The second parameter, \rprime,
corresponds to the derivative of the non-relativistic P-wave function
at the origin.
The third parameter is required to regulate the infrared divergency which
appears in the standard calculation of P-wave decays to a $g q \bar q$ final
state. This parameter is an infrared cutoff, usually loosely referred to as the
binding energy \ebin. It appears in the standard expression for the decay
widths  \cite{barbieri}\ as a factor $L=\log[4m_c^2/(4m_c^2-M^2)] =
\log(M/2\ebin)$, where \mc\ is the constituent charm quark mass. In
the recent formulation of P-wave decays introduced in ref. \cite{bbl1}, this
logarithm is absorbed into  the color octet, $^3S_1$ component of the $c\bar c$
wave function, which turns out to be proportional to the combination
$\as\rprime L$. As discussed in \cite{bbl2}, this relation provides a solid
basis for the study of higher order perturbative  corrections, as well as
providing a rigorous framework to regulate IR divergencies appearing in the
evaluation of the $q \bar q \to g \chi_J$ cross section.

We shall start by collecting here the expressions for NLO $\chi$ decay widths
\cite{barbieri}\ that will be used in our fit:
\ba
\label{ww1}
\chijgg &=&  A_J^{\gamma\gamma} e_Q^4 \aemsq  \frac{\rprime}{\mcfour }
    \left(1+ B_J \asopi \right)  \quad (J=0,2)
\ret
\label{ww2}
\chijlh &=&  A_J^{gg} \assq  \frac{\rprime}{\mcfour }
    \left(1+ C_J \asopi \right) +
   \pi \assq  H_8 \quad (J=0,2)
\ret
\label{ww3}
\chiolh &=&
   -\frac{56 n_f}{27} \assq  \frac{\as}{\pi} \frac{\rprime}{\mcfour } +
   \pi \assq  H_8
\ea
where:
\ba
A^{\gamma\gamma}_0 = 27 & A^{\gamma\gamma}_2 = \frac{36}{5} &
B_0 = \frac{\pi^2}{3}-\frac{28}{9} \quad
B_2 = -\frac{16}{3} \\
A^{gg}_0 = 6 & A^{gg}_2 = \frac{8}{5} &
C_0 = 8.772 \quad C_2 = -4.827 \quad n_f=3
\ea
To follow the spirit Ref. \cite{bbl1}, we introduced the
parameter \height, which we {\em define} as:
\be
\label{h8def}
	H_8 \= \frac{8 n_f}{9\pi} \frac{\as}{\pi}  \frac{\rprime}{\mcfour } L
\ee
The precise connection between this parameter and the color octet wave
function, can be found in \cite{bbl1}.

Following the suggestion of \cite{bbl2}, we chose to scale the widths by the
charm constituent quark mass, \mc, rather than by the mass of the quarkonium
states. This is consistent with the neglect of higher order non-relativistic
(NR) corrections, and with the inclusion of all non-perturbative effects into
\rprime. Since \rprime\ and $m_c$ always appear in the fixed combination
$\rprime/\mcfour $, the precise value of the charm mass will not change the
fitted
value of the other two free parameters, \as\ and \height.
It is important to point out, furthermore, that it is possible to make
consistent use of the extracted value of
\rprime\ in the calculation of production cross sections, by properly
identifying the origin of the mass terms present in the theoretical cross
sections. For example, LO direct $P$-wave production cross sections are
proportional to \rprime/$M^7$.
Of the seven powers of mass, four have the same origin as those appearing in
the
decay widths. The remaining three are from phase space. It is therefore
consistent to write the overall factor appearing in production as
\rprime/16\mcfour/$M^3$, where $M$ is the charmonium state mass. With this
choice, no ambiguity in the choice of \mc\ arises when using our fitted values
in the study of production cross sections.

In our expressions we neglected the very small contribution of the 3 gluon
decay of the \chio \cite{schuler}.
We included however a finite, non-logarithmic contribution to \chiolh, which
was not evaluated in the original calculation \cite{barbieri}.
To our knowledge this piece has always been neglected in previous studies
\footnote{We thank E. Braaten for pointing out this fact and suggesting we
evaluate and include  this contribution.}. It turns out that this term is not
numerically negligible when compared to the formally dominant logarithm, and
should therefore be included.

Corrections to the NR approximation are
not known, but are expected to at least partly cancel when taking ratios of
widths.  In a recent series of papers \cite{consoli},
Consoli and Field argued that by properly taking
into account the phase space reduction due to the effective mass of the gluon,
it is possible to provide a consistent perturbative description of the decay
widths of charmonium and bottomonium S-wave states,
without need for the inclusion of significant NR corrections.
This is consistent with the observation that EM decays
(such as $\eta_c \to \gamma \gamma$ or $\psi\to \ell^+\ell^-$) are correctly
predicted if the $O(v^2)$ corrections are set to 0. It is not known whether
this applies to $P$-waves as well, but we will comment on the
consequences of these ideas for our fits of $\chi$ decays at the end.

Taking ratios or differences of appropriate widths, can lead to expressions
which are independent of \rprime\ or \height, or both. These ratios can be used
to estimate with smaller theoretical uncertainty the only really perturbative
parameter of the theory, namely \as. Some examples which are often used in the
literature are:
\ba
\label{w1}
\frac{\chizgg}{\chitgg} &=& \left( \frac{15}{4} \right) \;
      \frac{1+\asopi B_0}{1+\asopi B_2}  \ret
\label{w2}
\frac{\chizlh-\chiolh}{\chitlh-\chiolh} &=& \left( \frac{15}{4} \right) \;
      \frac{1+(8.772+28/27)\asopi}{1+(-4.827+35/9)\asopi}  \ret
\label{w3}
\frac{\chizgg}{\chitlh-\chiolh} &=& \left( \frac{135}{8} \right) \;
     e_Q^4 (\frac{\aem}{\as})^2
      \frac{1+\asopi B_0}{1+(-4.827+35/9)\asopi}  \ret
\label{w4}
\frac{\chitgg}{\chitlh-\chiolh} &=& \left( \frac{9}{2} \right) \;
      e_Q^4 (\frac{\aem}{\as})^2
      \frac{1+\asopi B_2}{1+(-4.827+35/9)\asopi}
\ea
Notice that the inclusion of the non-logarithmic contribution to \chiolh\ (the
factor 35/9 appearing in the above equations)
significantly reduces the \oacube\ corrections to the difference
\chitlh--\chiolh.

In reference \cite{bbl1}\ the value of \as\ was extracted from the measurement
of bottomonium decays, evolved down to the charm mass.
This implicitly assumes that \as\
has to be evaluated at the scale of the heavy quark mass, and led to a value
of $\as(\mc)$=0.25$\pm$0.02 at \mc=1.5 \GeV. We prefer instead to leave \as\ as
a free parameter, to be fit together with \rprime/\mcfour\ and \height. In fact
we believe that the safest and less restrictive assumption on the scale of \as\
is that it has to be same for all decay processes, therefore enabling us to use
the same value of \as\ regardless of $J$ and of the final state. This choice is
also free of ambiguities related to the actual value of \mc. While this
will not allow us to extract a value of $\Lambda_{QCD}$, it provides however a
less restrictive constraint on the comparison of data
with QCD. We will verify at the end that the fitted value of \as\ is in
reasonable agreement with what expected fixing the renormalization scale to be
of the order of the charm mass.

Another choice performed in \cite{bbl1} was to use only the leading order (LO)
version of equations \eqnum{w1}-\eqnum{w4}. This was justified by the fact that
the contributions to decay widths via the color-octet $^3S_1$ component of the
wave function are only known to LO.
However, higher order corrections to the terms proportional to \height\
do not depend on $J$, as they arise from the $S$-wave component of the wave
function \cite{bbl2}. We can absorb these universal higher order corrections
into
the parameter \height, so that their effect does not change the results of the
fit
for the variables \as\ and \rprime. We therefore feel that it is justified to
use
the full \oacube\ expressions given above.

We collect the data that will be used in our fit in Table~1.
We notice that a key measurements, namely \chizgg, was
never actually published \cite{xball}. The large error associated to it will
not weigh this measurement significantly in our fit, which is unfortunate since
the ratio of the photonic widths of the even $\chi$ states is a very good probe
of the theory.
The other debated item is \chitgg, whose central value differs
significantly among various experiments \cite{E760-gg,TPC2-gg,CLEO-gg}
(we do not quote previous older results, some of which are simply upper
limits). It is not our duty to judge on the value of the experimental data, so
we chose, contrary to other authors \cite{consoli}, to include them all and let
the associated errors drive the fit.

We provide here the results of the fit to these seven measurements,
using the theoretical widths provided in equations \eqnum{ww1}-\eqnum{ww3}:
\ba  \label{ourfit}
&&
  \as = 0.286 \pm 0.031 \quad
  \frac{\rprime}{16\mcfour}= 0.60\pm 0.10  \; \MeV \quad
  \height = 4.2 \pm 0.7   \; \MeV \ret
&&
  \hone = 13.7 \pm 2.3   \; \MeV \quad \chi_{fit}^2 = 7.1
\ea
The total $\chi^2$ of the fit is 7.1 for the four
degrees of freedom. The poor quality of the fit is mostly due to the
discrepancy between the different measurements of \chitgg.
We included, for reference, also the numerical value of \hone, defined in
\cite{bbl1}\ as:
\be
\label{h1def}
 	H_1 \= \frac{9}{2\pi} \frac{\rprime}{\mcfour }
\ee
We notice that the value of \as\ returned by the fit is consistent with
$\as(m_c)=0.30$, obtained at two loops using \mc=1.5 \GeV\ and the value of
$\Lambda_4^{2-loop}=235$ \MeV\ extracted from DIS data \cite{mrsa}.
The value of the derivative of the wave function is also consistent with
potential model calculations (see for example the recent update in ref.
\cite{quigg}).

In Table~2 we provide the distribution of differences between expected
and measured decay widths, relative to the experimental errors, for the seven
measurements considered. The theoretical values are obtained using the fitted
values of parameters.
It is interesting to see what happens if one of the two best measurements of
\chitgg\ is removed from the fit. Removing the CLEO data point gives as central
values for the fit :
\ba
&&
  \as = 0.298 \pm 0.034 \quad
  \frac{\rprime}{16\mcfour}= 0.55\pm 0.11  \; \MeV \quad
  \height = 3.9 \pm 0.7   \; \MeV
  \ret
&&
  \hone = 12.6 \pm 2.3   \; \MeV  \quad
  \chi^2_{fit} = 3.5
\ea
The central values have not changed significantly from the global fit, but the
$\chi^2$ is now consistent with the 3 remaining degrees of freedom. The CLEO
measurement in this case would be off by less than 2 sigma from the theoretical
expectation.

Removing the E760 data point gives as central values for the fit :
\ba
&&
  \as = 0.195 \pm 0.031 \quad
  \frac{\rprime}{16\mcfour}= 1.23\pm 0.3   \; \MeV \quad
  \quad
  \height = 7.7 \pm 2.1     \; \MeV \quad \ret
&&   \hone = 28.3 \pm 7.8   \; \MeV  \quad
  \chi^2_{fit} = 3.1
\ea
The central values have now moved significantly. The $\chi^2$  has
improved, thanks to the larger relative error quoted by CLEO for \chitgg.
Notice that in this case the value of \as\ extracted is
consistent with what determined by CLEO in their analysis of their measurement
(\as=0.219$\pm$0.127 \cite{CLEO-gg}).
The E760 measurement is this case would be off by about 6 sigma from the
theoretical expectation based on the values of parameters extracted from the
fit.

In Table~3 we present the comparison between the width ratios given in
equations \eqnum{w1}-\eqnum{w4}\ and the data. The experimental error bars are
quite large, due to the propagation of errors in ratios of differences. It is
likely that some of the systematics or statistical errors are correlated and
will cancel in these combinations, but we did not pursue this possibility in
absence of enough details on the experimental analyses.
The same results, derived by excluding either E760 or CLEO from the fit, are
also included in Table~3.
%Notice that we preferred to directly to the single data, rather than to the
%ratio of differences. This is because, not having control over possible error
%correlations in the different measurements, we would be forced to add errors
%%in
%quadrature. The relative errors would therefore significantly increase.

As a final exercise, we include in our analysis the effect of an effective
gluon mass on the final state phase space. Following ref. \cite{consoli}, we
applied a correction factor to the hadronic decay widths. Notice that
contrary to $^3S_1$ decays, where the final state involves three gluons at
LO, there are only two gluons at LO in $\chi_{0,2}$ decays.
As a consequence, the impact of this correction is less significant. It is not
clear to us what is the right procedure to extend this idea to
decays to $q \bar q g$ final states. Since these are dominated by the
soft gluon region, where the quarks carry most of the energy, and since this
domain is already screened by the IR cutoff, we chose not to include any
correction factor for these final states.
We collect here the results of the global fit, which are only meant to
be indicative of the possible size of these effects:
\ba     \label{cffit}
&&
  \as = 0.326 \pm 0.030 \quad
  \frac{\rprime}{16\mcfour}=0.69 \pm 0.11  \; \MeV \quad
  \height = 4.2 \pm 0.5     \; \MeV \quad \ret
&&
  \hone = 15.8 \pm 2.0   \; \MeV   \quad
  \chi^2_{fit} = 6.6
\ea
The most significant change is in the value of \as, as already pointed out in
ref. \cite{consoli}. Notice also a slight improvement in the quality of the
fit.
We also mention that a clear prediction of the Consoli and Field
approach is that the ratio of hadronic widths of $\chi_0$ and $\chi_2$ should
be insensitive to the effective gluon mass.
The experimental value of this ratio is $7.9\pm 3.3$. The results
of our fit from Eq.\eqnum{ourfit}\ yield a theoretical ratio of $5.4\pm 1.3$,
those from Eq.\eqnum{cffit}\ yield $6.4\pm 1.4$. Both results are in agreement
with the data, the second one being slightly better because of the larger value
of \as.

In conclusion, we find that current data on $^3P_J$ charmonium decays are well
consistent with NLO perturbative QCD. The error bars are still large
for more incisive tests of the theory, and leave room for deviations from
the naive NR approximation and for higher order perturbative corrections.
Nevertheless the agreement found is encouraging, and hopefully relieves
serious concerns raised in earlier works.
A reduction in the significant discrepancy found between the \chitgg\
widths measured by different experiments will be of fundamental importance to
guide the extraction of theoretical parameters from the data.
Likewise, a new measurement of \chizgg\ would be very helpful.
\\[0.3cm]

\noindent
{\bf Acknowledgements:} We are very grateful to E. Braaten and M. Consoli for
several discussions and precious comments to this work.

\vspace{3cm}
{\renewcommand{\arraystretch}{1.8}
\begin{table}
\begin{center}
\begin{tabular}{lcl} \hline\hline
 Process & $\Gamma$ (\MeV) & Reference
\\ \hline\hline
$\chizgg $& $(4.0 \pm 2.8) \times 10^{-3}$ & Crystal Ball \cite{xball} \\
\hline
$\chitgg $& $(0.321\pm 0.095) \times 10^{-3}$ & E760 \cite{E760-gg} \\ \hline
$\chitgg $& $(1.08 \pm  0.38) \times 10^{-3}$ & CLEO \cite{CLEO-gg} \\ \hline
$\chitgg $& $(3.4 \pm  1.9) \times 10^{-3}$ & TPC2$_{\gamma}$ \cite{TPC2-gg} \\
\hline
$\chizlh $& $13.5 \pm  5.4$ & Crystal Ball \cite{xball} \\ \hline
$\chiolh $& $0.64 \pm  0.10$ & E760 \cite{E760-tot} \\ \hline
$\chitlh $& $1.71 \pm  0.21$ & E760 \cite{E760-tot} \\ \hline
\end{tabular}
\caption{\label{totexp}
Most recent experimental results on $\chi_c$ decay widths. The errors were
obtained by combining in quadrature the statistical and systematic errors given
in the quoted references. The widths to light hadrons were obtained from total
widths by removing the contributions of known radiative decays.}
\end{center}
\end{table} }

{\renewcommand{\arraystretch}{1.8}
\begin{table}
\begin{center}
\begin{tabular}{lc} \hline\hline
 Process & Data--Theory/Error
\\ \hline\hline
$\chizgg $& --0.44  \\ \hline
$\chitgg $ (E760)& 0.55\\ \hline
$\chitgg $ (CLEO)& --1.8\\ \hline
$\chitgg $ (TPC2)& --1.6\\ \hline
$\chizlh $& --0.74\\ \hline
$\chiolh $&  --0.13\\ \hline
$\chitlh $&  0.29\\ \hline
\end{tabular}
\caption{
Fractional differences, relative to the  experimental errors, between data and
theory predictions after the global fit. }
\end{center}
\end{table} }

{\renewcommand{\arraystretch}{2.0}
\begin{table}
\begin{center}
\begin{tabular}{lllll} \hline\hline
 & & Theory & & Data
\\
 & Global & No E760 & No CLEO & \\
\hline\hline
$\frac{\chizgg}{\chitgg} $ &
$ 7.4\pm 0.8 $ & $ 5.7 \pm 0.45 $ & $ 7.7\pm 0.9 $
     & $12.5 \pm 9.5$ (E760) \\
& & &  & $ 3.7 \pm 2.9$ (CLEO) \\
& & &  & $ 1.2 \pm 1.0$ (TPC2)
\\ \hline
$\frac{\chizlh-\chiolh}{\chitlh-\chiolh}  $ &
$ 7.7\pm 0.5 $ & $  6.4\pm 0.4 $ & $ 7.9\pm 0.5 $
                         & $ 12.0 \pm 5.7$ \\ \\ \hline
$\frac{\chizgg\times 10^3}{\chitlh-\chiolh}$ &
$  2.4 \pm 0.5 $  & $  5.0 \pm 1.5 $  &
$  2.2 \pm 0.5 $
                        & $  3.7 \pm 2.7  $ \\ \\ \hline
$\frac{\chitgg\times 10^4}{\chitlh-\chiolh}$ &
$  3.3 \pm 1.0 $  & $  8.8 \pm 3.4 $  &
$  2.9 \pm 1.0 $  &
        $ 3.0 \pm 1.1 $ (E760) \\
& & & & $ 10\pm 4  $ (CLEO) \\
& & & & $ 32\pm 19  $ (TPC2) \\ \hline
\end{tabular}
\caption{Comparison between data and theory for ratios of widths
(includes global fit, fit without E760 and without CLEO \chitgg\ datum).
}
\end{center}
\end{table} }

\end{document}